\documentstyle[sprocl,epsfig]{article}

\newcommand{\alt}{\,\rlap{\lower4.5pt\hbox{$\mathchar\sim$}}\raise2.5pt
\hbox{$<$}\,}

\begin{document}

\title{\boldmath
THEORETICAL DEVELOPMENTS IN STANDARD-MODEL HIGGS PHYSICS AT A FUTURE
$e^+e^-$ LINEAR COLLIDER\footnote{%
To appear in the {\it Proceedings of the Workshop on Physics and Experiments
with Future Electron-Positron Linear Colliders (LCWS 2002)}, Jeju Island,
Korea, August 26--30, 2002.}
\unboldmath}

\author{B.A. KNIEHL}

\address{II. Institut f\"ur Theoretische Physik, Universit\"at Hamburg,
Luruper Chaussee 149, 22761 Hamburg, Germany}

\maketitle\abstracts{
We review the decay properties and production mechanisms of the Standard-Model
Higgs boson at a future $e^+e^-$ linear collider with special emphasis on the
influence of quantum corrections.
We also discuss how its quantum numbers and couplings can be extracted from 
the study of appropriate final states.}

\vspace*{-0.8cm}

\section{Introduction}

\vspace*{-0.2cm}

The Higgs boson is the missing link of the Standard Model (SM) of elementary
particle physics.
At a future $e^+e^-$ linear collider (LC), an important experimental task will
be to determine the Higgs quantum numbers and couplings in order to
distinguish between the minimal SM and possible extensions.
In particular, the measurement of the Higgs self-couplings will allow one to
directly test the Higgs mechanism.

In this presentation, we discuss the decay properties of the Higgs boson and
its main production mechanisms at a future $e^+e^-$ LC, emphasizing the
influence of radiative corrections, and explain how to extract its quantum
numbers and couplings from the study of final states.
For more details, we refer to a recent review.\cite{bak}

\vspace*{-0.4cm}

\section{Decay Properties}

\vspace*{-0.2cm}

At the tree level, the Higgs boson decays to pairs of massive fermions and
gauge bosons if $M_H>2M_i$ ($i=f,V$).
If $M_V<M_H<2M_V$ ($M_H<M_V$), then one of the (both) final-state gauge bosons
are forced to be off shell, so that one is dealing with three-particle
(four-particle) decays.\cite{hzgg}
The Higgs boson also couples to photons (gluons), through loops involving
charged (coloured) massive particles, and one is led to consider the
loop-induced decays $H\to Z\gamma$, $H\to\gamma\gamma$, $H\to gg$, {\it etc.}

In order to match the high experimental precision to be achieved with a future
$e^+e^-$ LC, it is indispensable to take radiative corrections into account.
At one loop, the electroweak corrections to
$\Gamma\left(H\to f\bar f\right)$,\cite{fle,hff}
$\Gamma(H\to VV)$,\cite{fle,hzz} and
$\Gamma\left(H\to Zf\bar f\right)$\cite{pr} and the QCD ones to
$\Gamma(H\to q\bar q)$\cite{hqq} are well established, including the
dependence on all particle masses.
Beyond one loop, only dominant classes of corrections were investigated,
sometimes only in limiting cases.
For a low-mass Higgs boson, these include corrections enhanced by the
strong-coupling constant $\alpha_s$ and the top Yukawa coupling
$g_{ttH}=M_t/v$, where $v=2^{-1/4}G_F^{-1/2}\approx246$~GeV, with $G_F$ being
Fermi's constant.
Specifically, the two-loop QCD corrections were found for
$\Gamma(H\to l^+l^-),$\cite{hll}
$\Gamma(H\to q\bar q)$ ($q\ne t$),\cite{gor,lar}
$\Gamma\left(H\to t\bar t\,\right),$\cite{har}
$\Gamma(H\to Z\gamma),$\cite{spi} $\Gamma(H\to\gamma\gamma),$\cite{zhe}
and $\Gamma(H\to gg)$.\cite{lar,ina}
Even three-loop QCD corrections were calculated, namely for
$\Gamma(H\to q\bar q)$ ($q\ne t$),\cite{che}
$\Gamma(H\to\gamma\gamma),$\cite{ste} and $\Gamma(H\to gg)$.\cite{cks}
In the last case, they are quite significant, the correction factor
being $1+0.66+0.21$ for $M_H=100$~GeV.\cite{cks}

An efficient way of obtaining corrections leading in
$X_t=g_{ttH}^2/(4\pi)^2$ to processes involving low-mass Higgs bosons is to
construct an effective Lagrangian by integrating out the top quark.
This may be conveniently achieved by means of a low-energy theorem,\cite{let}
which relates the amplitudes of two processes which differ by the insertion of
an external Higgs-boson line carrying zero four-momentum.
In this way, the $O\left(X_t^2\right)$,\cite{gam}
$O(X_t\alpha_s)$,\cite{hll,was} and $O\left(X_t\alpha_s^2\right)$\cite{fer}
corrections to $\Gamma(H\to l^+l^-)$, $\Gamma(H\to W^+W^-)$, and
$\Gamma(H\to ZZ)$ were obtained.
In $O\left(X_t^2\right)$, also the full $M_b$ dependence is
available.\cite{gam}
The $O(X_t)$, $O\left(X_t^2\right)$, and $O(X_t\alpha_s)$ corrections to
$\Gamma(H\to q\bar q)$, where $q\ne b,t$, coincide with those for
$\Gamma(H\to l^+l^-)$.
The $O\left(X_t\alpha_s^2\right)$ corrections to $\Gamma(H\to q\bar q)$ were
found in Ref.~22.
The effective-Lagrangian method in connection with the low-energy theorem was
also employed to obtain the $O(X_t\alpha_s)$\cite{hbb} and
$O\left(X_t\alpha_s^2\right)$\cite{cks1} corrections to
$\Gamma(H\to b\bar b)$, the $O(X_t)$ corrections to
$\Gamma(H\to\gamma\gamma),$\cite{gam} and the $O(X_t)$\cite{gam,cks1,dg} and
$O(X_t\alpha_s)$\cite{mat} corrections to $\Gamma(H\to gg)$.

It is fair to say that radiative corrections for Higgs-boson decays have been
explored to a similar degree as those for $Z$-boson decays.
Unfortunately, this does not necessarily lead to similarly precise theoretical
predictions.
In fact, the errors on the latter are dominated by parametric uncertainties,
mainly by those in $\alpha_s^{(5)}(M_Z)$ and the quark masses.\cite{gro}

\vspace*{-0.4cm}

\boldmath
\section{Production in $e^+e^-$ Collisions}
\unboldmath

\vspace*{-0.2cm}

The dominant mechanisms of Higgs-boson production in $e^+e^-$ collisions are
Higgs-strah\-lung and $W^+W^-$ fusion.
The cross section of $ZZ$ fusion is approximately one order of magnitude
smaller than the one of $W^+W^-$ fusion, because of weaker couplings.
Compact cross section formulas may be found in Ref.~27.

As for the Higgs-strahlung process, the electromagnetic\cite{ffzh,ber} and
weak\cite{ffzh,den} corrections are fully known at one loop.
The electroweak corrections for $VV$ fusion, a $2\to3$ process, are not yet
available.
However, the leading effects can be conveniently included as follows.
The bulk of the initial-state bremsstrahlung can be taken into account in the
so-called leading logarithmic approximation provided by the structure-function
method, by convoluting the tree-level cross section with a radiator function,
which is known through $O(\alpha^2)$ and can be further improved by
soft-photon exponentiation.\cite{bee}
The residual dominant corrections of fermionic origin can be incorporated in a
systematic and convenient fashion by invoking the so-called improved Born
approximation.\cite{was,fer,iba}

\vspace*{-0.4cm}

\section{Quantum Numbers and Couplings from Final States}

\vspace*{-0.2cm}

The spin, parity, and charge-conjugation quantum numbers $J^{PC}$ of Higgs
bosons can be determined at a future $e^+e^-$ LC in a model-independent way
by analyzing the threshold behaviour and the angular dependence of the
Higgs-strahlung process.\cite{ver}
This process can also be employed to place limits on anomalous $ZZH$ and
$Z\gamma H$ couplings.\cite{hag}
The top Yukawa coupling $g_{ttH}$ and the trilinear Higgs self-coupling
$g_{HHH}$ can be probed by studying the processes
$e^+e^-\to t\bar tH$,\cite{gun} $e^+e^-\to ZHH$, and
$e^+e^-\to\bar\nu_e\nu_eHH$.\cite{mmm} 

\vspace*{-0.4cm}

\section{Conclusions and Outlook}

\vspace*{-0.2cm}

The theoretical predictions for the partial decay widths of the SM Higgs boson
and its production cross sections at a future $e^+e^-$ LC are generally in
good shape.
The strategies for the determination of the Higgs profile are also well
elaborated.
The list of urgent tasks left to be done includes the calculation of the full
$O(\alpha)$ corrections for important $2\to3$ processes, such as $W^+W^-$ 
fusion, $ZZ$ fusion, and $t\bar tH$ associated production, and the inclusion
of background processes and detector simulation.

\end{document}